\pdfoutput=1
\documentclass[10pt]{jpaper}

\usepackage{etex}
\usepackage[sort,nocompress,adjust]{cite} % autosort groups of citations
\usepackage{amsmath,amssymb,amsfonts}
\usepackage{algorithmic}
\usepackage{graphicx}
\usepackage{listings}
\usepackage{xcolor}
\definecolor{olive}{rgb}{0.33, 0.42, 0.18}
\usepackage{makecell}
\usepackage{textcomp}
\usepackage{fancyhdr}
\usepackage{tabu}
\usepackage{enumitem}
\usepackage[normalem]{ulem}
\usepackage{inconsolata}
\usepackage[scaled]{beramono}
\usepackage{float}
\usepackage{caption}

\def\BibTeX{{\rm B\kern-.05em{\sc i\kern-.025em b}\kern-.08em
T\kern-.1667em\lower.7ex\hbox{E}\kern-.125emX}}

\usepackage{verbatim}   % allows multi-line comments

\usepackage{wrapfig}
\usepackage{multirow}
\usepackage{balance}
\usepackage{cite}

\usepackage{pifont}

\usepackage{rotating}

\usepackage{arydshln}

\usepackage{placeins}

\usepackage{gensymb}

\usepackage{enumitem}
\usepackage[ruled,vlined]{algorithm2e}
\usepackage{upgreek,textgreek}
\usepackage[htt]{hyphenat}
\usepackage{appendix}

\hypersetup{colorlinks,
	linkcolor=blue,
	citecolor=blue,
	urlcolor=blue,
	filecolor=blue}

\usepackage{mathtools}

\usepackage{tikz}

\usepackage{listings}
\usepackage[utf8]{inputenc}
\lstdefinestyle{customcpp}{
	 aboveskip=0in,
	  belowskip=0in,
	   abovecaptionskip=0in,
	    belowcaptionskip=0in,
	     captionpos=b,
	      xleftmargin=\parindent,
	       language=C++,
	        morekeywords={forall},
		 showstringspaces=false,
		  basicstyle={\linespread{0.6}\fontseries{sb}\small\ttfamily},
		   keywordstyle=\bfseries,
		    commentstyle=\itshape\color{green!40!black},
	    }

\begin{document}

%
% The "title" command has an optional parameter, allowing the author to define a "short title" to be used in page headers.
%\title{Sage: Practical \& Scalable ML-Driven Performance Debugging in Microservices\vspace{-0.16in}}
\title{Sinan: Data-Driven Resource Management for Cloud Microservices\vspace{-0.16in}}
%
% The "author" command and its associated commands are used to define the authors and their affiliations.
% Of note is the shared affiliation of the first two authors, and the "authornote" and "authornotemark" commands
% used to denote shared contribution to the research.
\author{Yanqi Zhang, Weizhe Hua, Zhuangzhuang Zhou, Edward Suh, and Christina Delimitrou\\ Cornell University\\Contact author: yz2297@cornell.edu}

\date{}

\maketitle
\pagestyle{plain}

\begin{abstract}
Cloud applications are increasingly shifting to interactive and loosely-coupled microservices. Despite their advantages, 
microservices complicate resource management, due to inter-tier dependencies. 
We present Sinan, a cluster manager for interactive microservices that leverages easily-obtainable tracing data instead of empirical decisions, to infer the impact of a resource allocation on on end-to-end performance, and allocate appropriate resources to each tier. 
In a preliminary evaluation of Sinan with an end-to-end social network built with microservices, we show that Sinan's data-driven 
approach, allows the service to always meet its QoS without sacrificing resource efficiency. 

\end{abstract}

\section{Introduction}
Over the past few years, the design of online interactive applications has shifted from \textit{monolithic} services that encompass the entire functionality in a single binary, to \textit{microservices} that divide application to a graph of tens or hundreds of single-purpose and loosely-coupled tiers
~\cite{suresh2017distributed,gan2019open,sriraman2018mu,Cockroft15,Cockroft16,twitter_decomposing}. 
%\begin{wrapfigure}[13]{l}{0.28\textwidth}
%	\centering
%	\includegraphics[scale=0.28, trim=0 0.2cm 0 5.4cm, clip=true]{figures/motivation_comparison3.pdf}
%	\caption{\label{fig:motivation} \figurecaptionsize{Differences in the design and deployment of monoliths and microservices.} }
%\end{wrapfigure}
\textit{Microservices} are appealing for several reasons, including modularity, flexible development and deployment, and high tolerance of software heterogeneity.

Despite their advantages, microservices also introduce new system challenges, primarily in resource management. The dependencies between microservices exacerbate queueing and introduce cascading QoS violations that are difficult to identify and correct in a timely manner~\cite{gan2019open,zhou2018overload}. Current cluster managers 
that optimize for monolithic applications or applications consisting of few tiers are not expressive enough to capture the complexity of microservices\cite{Delimitrou13,Delimitrou14,Delimitrou13d,Delimitrou14b,Delimitrou15,gan2018seer,Delimitrou16,Delimitrou17,Delimitrou19,
Borg,Lin11,Meisner11,Lo14,Ousterhout13,omega13,Cloudscale}. 
Furthermore, given that typical microservice deployments include tens to hundreds of unique tiers, exhaustively exploring the resource allocation space is prohibitively expensive~\cite{Lo15,Delimitrou14}. 

Instead in this work we take a data-driven approach to microservices management. Previous work~\cite{cortez2017resource,rzadca2020autopilot} highlighted the potential of data-driven approaches to address resource scheduling for large-scale systems, but they do not directly apply to microservices.

We present our preliminary work on Sinan, an ML-based cluster manager for microservices that leverages the cloud's tracing data and a set of practical ML techniques to infer the impact of resource allocation on end-to-end performance, 
and assign appropriate resources to each application tier. Sinan leverages efficient action space pruning to reduce the overheads of exploration, and trains two models with the tracing data; a CNN model for detailed short-term performance prediction, and, a Boosted Trees model that evaluates the long-term performance evolution. The combination of the two models allows Sinan to both examine the near-future outcome of a resource allocation, and account for the system's inertia in building up queues. Sinan operates online, adjusting per-tier resources dynamically according to the service's runtime status and end-to-end Quality of Service (QoS) target.

We evaluate Sinan using an end-to-end, microservices-based application that implements a social network~\cite{gan2019open}, and compare it against traditional autoscaling approaches. 
We also validate the accuracy of Sinan's ML models, and show that QoS does not come at the price of resource inefficiency. 
Finally, we emphasize the need for explainable ML models, which provide design insights for large-scale systems, using an example of Redis's log synchronization, which Sinan identified as the source of unpredictable performance. 
% Systems like Sinan show that ML-driven approaches 
% offer practical solutions for systems whose scale make previous empirical approaches impractical. 

\section{Overview}
\label{sec:overview}

\subsection{Motivating Application}
\label{sec:applications}
\begin{figure}[h!]
\centering
  \includegraphics[scale=0.26,viewport=260 30 700 400]{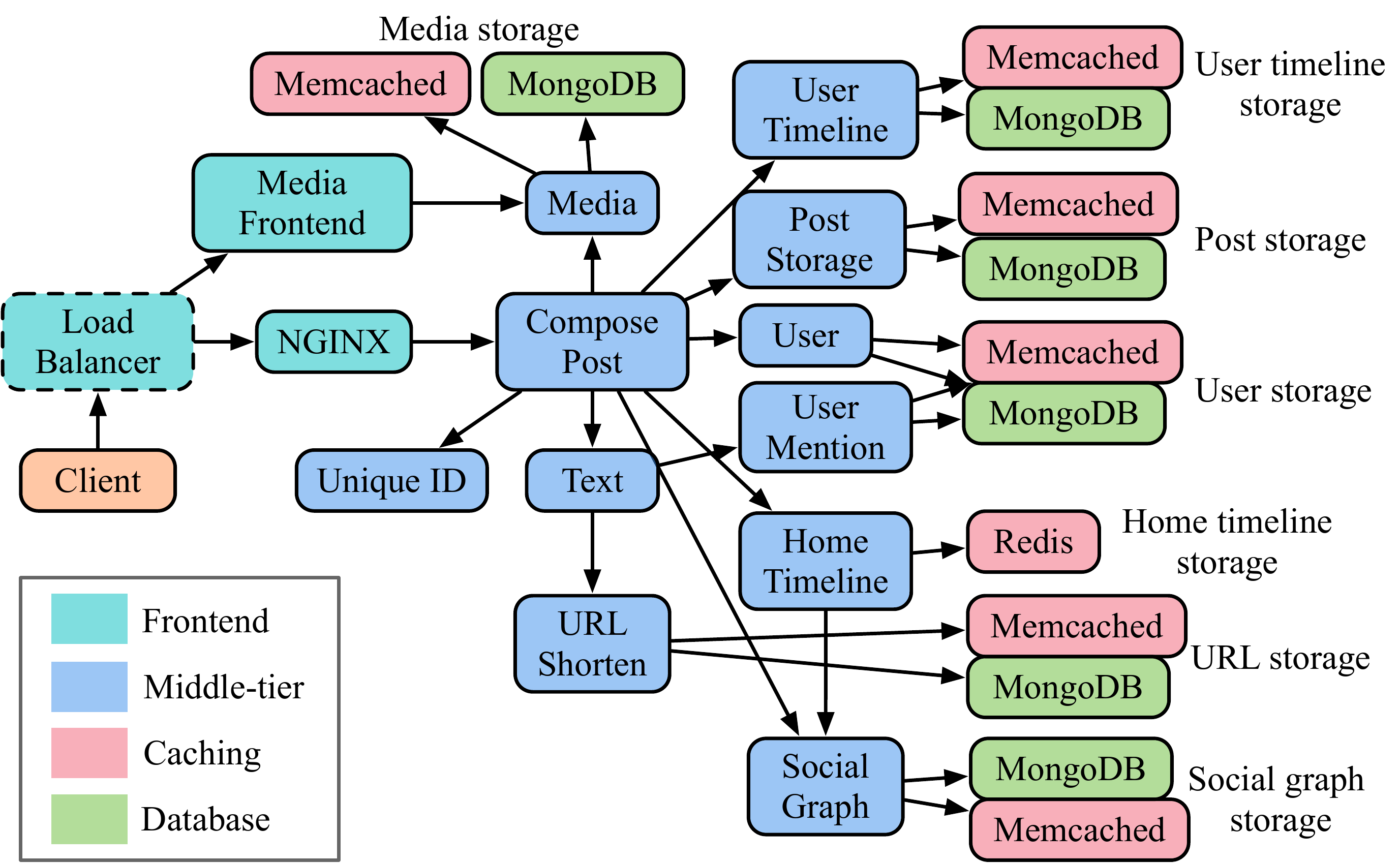}
  \caption{Social Network: vertical lines separate different types of tiers (front-end, logic, backend), 
  while colors show microservices for different request types. Blue: read own timeline, green: read user timeline, orange: compose post.  }
  \label{fig:social_network}
  \vspace{-0.08in}
\end{figure}

We use the \texttt{Social Network} from DeathStarBench~\cite{gan2019open}. 
The service implements a broadcast-style social network with uni-directional follow relationships, and its architecture is shown in Fig.~\ref{fig:social_network}. 

%Block size represents the initial over-provisioned core allocation per tier.
\noindent{\bf{Functionality: }}
%The blue rectangles in each microservice signify tracing probe points. 
Users ({{\texttt{client}}}) %on the service
send requests over {\texttt{http}} to {\texttt{NGINX}} front-end, which selects a specific downstream service to forward the request to.
Users can create posts embedded with text, media, links, and tags to other users, which are then broadcasted to all their followers. Users can also read posts on their timelines and create new posts themselves. 
 
Inter-microservice messages use Thrift RPCs~\cite{thrift}. The service's backend uses {\texttt{memcached}} for caching, and {\texttt{MongoDB}} for persistently storing posts, user profiles,
and media. Index information, such as user timeline indices, are stored in {\texttt{Redis}}.  {\texttt{RabbitMQ}} instances are added between business logic and {\texttt{MongoDB}} to make time-consuming database write operations asynchronous, and prevent them from blocking upstream connections.  
We use the {\texttt{Reed98}}~\cite{nr} social friendship network to populate the user database. It is extracted from Facebook, and consists of 962 people and 18.1K edges, where an edge is a follow relationship. We model user activity according to the behavior of Twitter users reported in~\cite{kwak2010twitter}, where a user's posting activity positively correlates with the number of their followers. The distribution of post length emulates Twitter's text length distribution~\cite{gligoric2018constraints}. 
%In our \texttt{Social Network} deployment, we assume that the social friendship graph is rarely changed and focus on measuring performance of reading and composing posts. 
\vspace{-0.1in}

\subsection{Management Challenges and the Need for ML}
\label{sec:challenges}

Microservices management faces three major challenges. 

\noindent\textbf{1. Prohibitively-large action space} Given that application behaviors change frequently, 
resource management decisions need to happen online. This means that the resource manager must traverse 
a space that includes all possible resource allocations per microservice in a practical manner. 
Assuming $N$ microservice tiers and a pool of $C$ ($C \geq N$) homogeneous physical cores, each with $F$ frequency levels, 
the size of the action space is $\binom{C-1}{N-1} \cdot N^F$. For example, given a cluster with 150 cores and assuming 10 frequency steps per tier, the resource allocation space size of the \textit{Social Network} application is $7.78\times10^{55}$. Profiling all actions under different loads would require significant time and computation resources. As a result, efficient action space pruning methods and statistical tools with strong generalization capability are urgently needed for resource scheduling.

\noindent\textbf{2. Delayed queueing effect} Consider a queueing system with processing throughput $T_o$ under a latency QoS target. 
$T_o$ is a non-decreasing function of the amount of allocated resources $R$. For input load $T_i$, $T_o$ should equal or slightly 
surpass $T_i$ for the system to meet its QoS, and remain stable, while using the minimum amount of resources $R$ needed. 
Even in the case where $R$ is reduced, such that $T_o < T_i$, QoS will not be immediately violated, since queue accumulation takes time. 
The converse is also true; by the time QoS is violated, the built-up queue takes a long time to drain, even if resources are upscaled immediately upon detecting the QoS violation. Multi-tier microservices are complex queueing systems with queues both between and within microservices~\cite{gan2019open,gan2018seer,Delimitrou19}. This delayed queueing effect highlights the need for the ML model to evaluate the long-term effect of resource management actions, and proactively prevent the resource manager from reducing resources overly-aggressively to avoid latency spikes that introduce long recovery periods. To mitigate a QoS violation, % caused by deallocating resources, 
the manager must increase resources proactively, otherwise the QoS violation becomes unavoidable, even if more resources are allocated a posteriori.

%The complexity and scale of microservices-based applications make resource management challenging, especially given the fact that dependent tiers are not perfect pipelines, and hence can introduce backpressure effects that are hard to detect and prevent~\cite{Delimitrou19,Gan19,gan2018seer}.

\noindent{\bf{3. Dependencies among tiers}} 
Resource management in microservices is additionally complicated by the fact that dependent microservices are not perfect pipelines, and hence can introduce backpressure effects that are hard to detect and prevent~\cite{gan2019open,gan2018seer,Delimitrou19}. %Dependencies often exist among individual microservices, as multiple tiers come together to implement the end-to-end service. 
These dependencies can be further exacerbated by the specific RPC and data store API implementation. 
%Dependencies complicate identifying bottleneck tiers, as shown by the backpressure effect described in~\cite{gan2019open}. 
Therefore, the resource scheduler should have a global view of the microservice graph and be able to anticipate the impact of dependencies on end-to-end performance.

%+++ add description and figure. 

\subsection{Proposed Approach}

These challenges suggest that empirical resource management, such as autoscaling, is prone to unpredictable performance and/or resource inefficiencies. Sinan takes instead a data-driven, ML-based approach that automates resource management for microservices, leveraging a set of scalable and validated ML models, that allows high and predictable performance and resource utilization. 
%In order to handle the complexity of microservices, Sinan uses machine learning models to predict the performance of an application for potential resource configurations.
Sinan's ML models predict the end-to-end latency and the probability of a QoS violation for a resource configuration, given the system's state and history. The system uses these predictions to maximize resource efficiency, while meeting the application's QoS. Below, we first describe the ML models (Sec.~\ref{sec:design_models}), and Sinan's system architecture (Sec.~\ref{sec:design}).

\section{Sinan}
\label{sec:design}

\subsection{Machine Learning Models}
\label{sec:design_models}

We initially designed Sinan's ML model to only predict the end-to-end latency tail distribution using a CNN, such that the scheduler can query the model with potential resource allocations, and 1) choose the one that minimizes the required resources while meeting the end-to-end QoS, or 2) the one that minimizes end-to-end latency if there are multiple allocations satisfying QoS. However, this model experienced consistently high errors during deployment, due to the delayed queueing effect mentioned previously. Therefore, it is crucial for the model to also predict the long-term impact of resource allocations.  

A straightforward fix is to use a multi-task CNN model that predicts latency distribution for the immediate future, and the QoS violation probability in the long term. This approach was also shown to frequently overpredict latency, due to the large gap between latency and probability values. 

Sinan currently follows a hybrid approach, which uses a CNN to predict the end-to-end latency of the next decision interval, and a boosted trees (BT) model to anticipate QoS violations further into the future. 
We refer to the CNN model as the \textit{short-term latency predictor}, and the BT model as the \textit{long-term violation predictor}.

As shown in Fig.~\ref{fig:models}, the latency predictor takes as input the resource usage history ($X_{RH}$), 
the latency history ($X_{LH}$), and the potential resource configuration ($X_{RC}$) for the next timestep, 
and predicts the end-to-end tail latency distribution ($y_L$) (95$^{th}$ to 99$^{th}$ percentiles) of the next timestep.
$X_{RH}$ is formed as a 3D tensor whose x-axis is $N$ tiers in the microservices graph, y-axis is $T$ timestamps ($T>1$ accounts for the non-Markovian nature of microservice graph), and channels are $F$ resource usage information related to cpu and memory. $X_{RC}$ and $X_{LH}$ are 2D matrices. For $X_{RC}$, x-axis is $N$ tiers and y-axis is core number and cpu frequency. For $X_{RH}$,  x-axis is $T$ timestamps, and y-axises are latency tail distribution (95$^{th}$ to 99$^{th}$ percentiles). The three inputs are firstly individually processed with convolution (Conv) and fully connected (FC) layers and then concatenated to form the latent representation $L_{f}$, from which the predicted latency distribution $L_{f}$ is derived with another FC layer. The loss function of the CNN is shown below:

\begin{figure}[t]
    \centering 
    \includegraphics[scale=0.13]{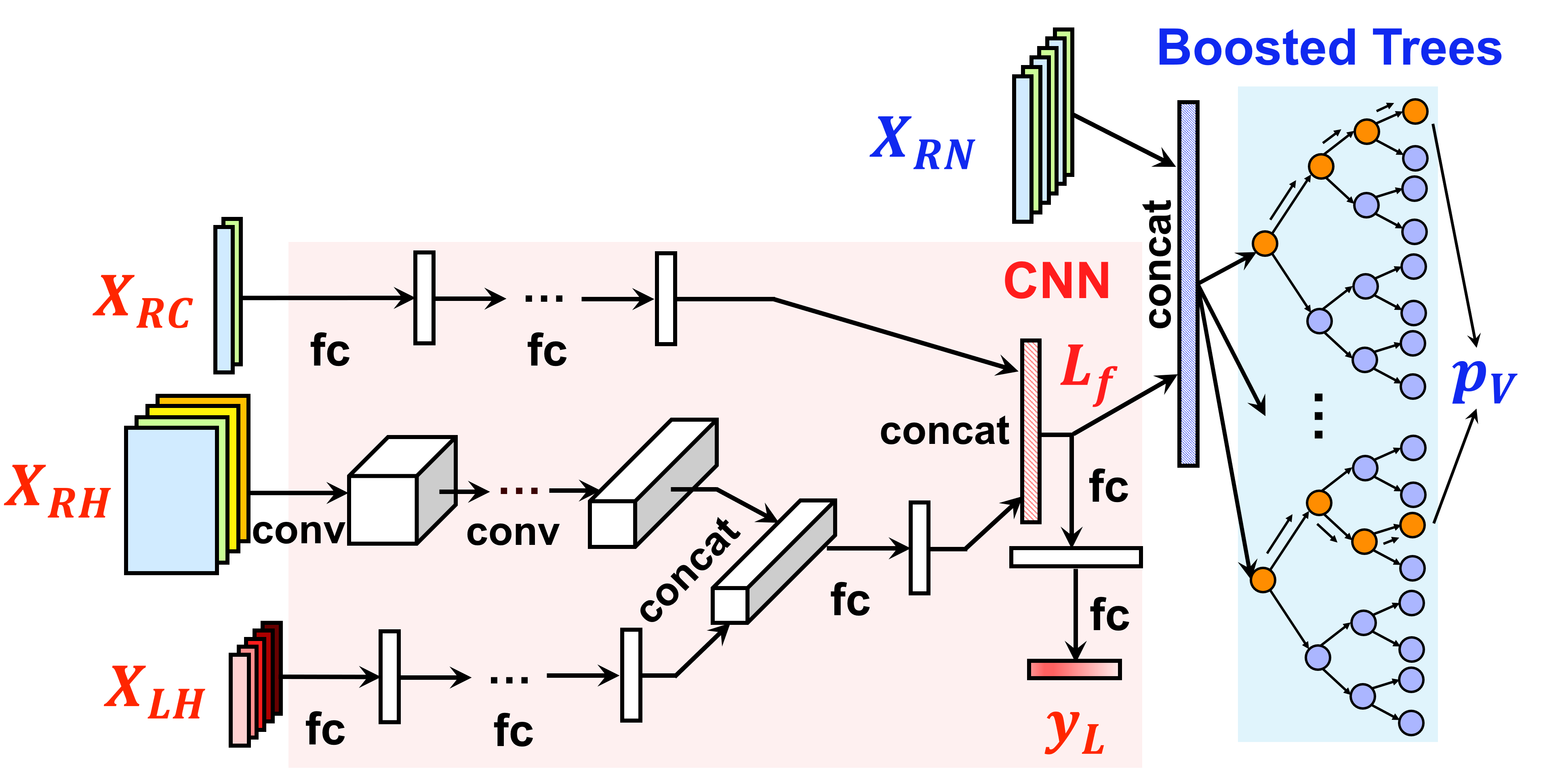}
    \caption{Sinan's hybrid model, consisting of a CNN and a boosted trees model. The CNN predicts the end-to-end latency ($y_L$), 
	    and the boosted trees predicts the probability of a QoS violation ($p_V$).}
    \label{fig:models}
    \vspace{-0.14in}
\end{figure}

%\textbf{Custom Loss Function} \hspace{3pt}
\vspace{-0.08in}
\begin{equation}
    \mathcal{L}(X, \hat{y}, W) = \frac{1}{n}\sum_i^n (\hat{y_i}-f_W(x_i))^2
    \label{eq:square_loss}
\end{equation}
\vspace{-0.04in}
where $f_W(\cdot)$ represents the forward function of the CNN network, $\hat{y}$ is the ground truth, and $n$ is the number of training samples. 
%Given the spiking behavior of interactive microservices 
%that leads to very high latency, the squared loss in Eq.~\ref{eq:square_loss} tends to overfit for training samples 
%with large end-to-end latency, which have large absolute squared loss. As a result, the model always overpredicts latency. 
%Since the latency predictor aims to find the best resource allocation subject to a QoS target, 
%the loss should be biased towards training samples whose end-to-end latencies are $\leq QoS$. 
%Therefore, we use a scaling function to scale both the predicted and actual end-to-end latency 
%before applying the squared loss function. The scaling function ($\phi(\cdot)$) is defined as:
%\begin{equation}
%\phi(x)=
%\begin{cases}
%    x& \text{$x$} \leq \text{$t$}\\
%    t + \frac{x-t}{1+\alpha(x-t)} & \text{$x > t$}
%\end{cases}
%\label{eq:1}
%\end{equation}
%\vspace{-0.02in}
%where the latency range is in $(0,t)$ and $\alpha$ is a hyper-parameter which can be tuned to achieve different decay effects. 
%%The scaling factor helps generalize the model to non-problematic cases, when the system is not saturated to avoid biases from the few QoS violations. 
%As shown in Fig.~\ref{fig:loss_func}, we visualize the scaling function with $t = 100$ and $\alpha = 0.005, 0.01, 0.02$, respectively. 
%{$\alpha$ is empirically selected for one service, and validated against the remaining. }
%The custom loss function is then formulated as:
%\begin{equation}
%  \mathcal{L}(X, \hat{y},W)=\sum_i^n(\phi(\hat{y_i})-\phi(f_W(x_i)))^2
%  \label{eq:custom_loss}
%\end{equation}
%\vspace{-0.04in}
The CNN is implemented with MxNet~\cite{mxnet}, and trained with Stochastic Gradient Descent (SGD). 
%a high performance deep learning framework. All models are trained using Stochastic Gradient Descent (SGD).
%\vspace{0.04in}

%\begin{minipage}{\linewidth}
%  \begin{minipage}{0.51\linewidth}
%    \begin{tabular}{cp{3.5cm}}
%    \footnotesize
%    \textbf{{\footnotesize {Param.}}} & \textbf{\footnotesize{{Definition}}}\\
%    \textbf{\footnotesize{{k}}} & {\footnotesize{{future timesteps in BT}}} \\[0.06cm]
%    \textbf{\footnotesize{{T}}} & {\footnotesize{{past timesteps in CNN\&BT}}} \\[0.06cm]
%    \textbf{\footnotesize{{N}}} & {\footnotesize{{application tiers}}} \\[0.06cm]
%    \textbf{\footnotesize{{M}}} & {\footnotesize{{latency percentiles}}} \\[0.06cm]
%    \textbf{\footnotesize{{F}}} & {\footnotesize{{resource statistics}}} \\[0.06cm]
%    \textbf{\footnotesize{{R}}} & {\footnotesize{{allocated resources}}} \\[0.06cm]
%\end{tabular}
%\captionof{table}{{Parameters for the hybrid ML model. }}
%    \end{minipage}
%    \hspace{0.1cm}
%  \begin{minipage}{0.41\linewidth}
%    \centering
%    \includegraphics[scale=0.24,viewport=20 0 400 260]{figs/scale.pdf}
%    \vspace{-0.25cm}
%    \captionof{figure}{\label{fig:loss_func}Scale function $\phi(\cdot)$ with different $k$.}
%  \end{minipage}
%  \end{minipage}
%
%\vspace{-0.08in}
%\subsection{Violation Predictor}
%\vspace{-0.06in}

The violation predictor %, on the other hand, 
addresses a binary classification task of whether a given allocation will cause a QoS violation further in the future, to filter out undesirable resource options. Ensemble methods are good candidates for this, as they usually perform well in classification tasks, and are robust to overfitting. We use XGBoost~\cite{boosting}, 
which realizes an accurate non-linear model by combining a series of simple regression trees. 
It models the target as the sum of trees, each of which maps the features to a score. The final prediction is the accumulation of scores across all trees. We use the compact latent variable $L_f$, extracted from the CNN as the input to the BT, to reduce the computation cost. Moreover, since the latent variable $L_f$ is significantly smaller than $X_{RC}$, $X_{RH}$, and $X_{LH}$ in dimensionality, using $L_f$ as the input also makes the model more robust to overfitting. 

%Boosted Trees also take resource allocations as input. During inference, we simply use 
%the same resource configuration for the next $k$ time steps to predict whether it will 
%cause a QoS violation $k$ steps into the future. As illustrated in Fig.~\ref{fig:models}, 
%each leaf of the trees represents either a violation or non-violation with a continuous score. 
%For a given example, we sum the scores for all chosen violation leaves ($s_{V}$) and non-violation 
%leaves ($s_{V}$) from each tree. The output of the boosted tree is the predicted probability of QoS violation ($p_{V}$), 
%which can be calculated as $p_{V} = \frac{e^{s_{V}}}{e^{s_{V}}+e^{s_{NV}}}$. For the violation predictor  
%we leverage XGBoost~\cite{xgboost}, a gradient tree boosting framework that improves scalability using sparsity-aware 
%approximate split finding, to build and train our model. 

\vspace{-0.06in}

\subsection{Online Scheduler}
\label{sec:design_overview}

Sinan consists of three components: a centralized scheduler, distributed operators deployed on each server, and a performance forecaster hosting the ML models.%, as shown in Fig.~\ref{fig:sys}. 

% \begin{figure}
% \centering
%   \includegraphics[scale=0.16,viewport=450 50 1000 450]{figs/sys.png}
%   \caption{Sinan's system architecture.}
%   \label{fig:sys}
%   \vspace{-0.14in}
% \end{figure}

Sinan makes decisions periodically, once every 1s, consistent with the granularity at which QoS is defined. 
% (determined based on the granularity at which QoS is defined), 
The centralized scheduler queries the distributed operators to obtain the CPU, memory, network, and I/O usage information of each tier in the previous interval through Docker's cgroup stats API. Aside from per-tier information, the scheduler also queries the API gateway to get statistics of user load (implemented via workload generator for simplicity in our experiments). %Once the centralized scheduler receives this data, 
%it sends it to the hybrid ML model, which is responsible for evaluating the impact of different resource allocation decisions. 
%Decisions are limited to a predefined set of operations to minimize scheduling overheads. %enable timely evaluation and speed up training data collection.
Using the model's output, the scheduler chooses one allocation that is beneficial to QoS and resource-efficient, i.e., uses the least resources needed to meet QoS, and sends its decision to the per-node agents for enforcement. %which are responsible for changing core allocation and frequency setting. 

%\subsubsection{Implementation}impl
\subsubsection{Data Collection}
\label{sec:data_collection}

Representative training data is key to the accuracy of ML models. %, since algorithms like Stochastic Gradient Descent (SGD) cannot provide any guarantees on the inference accuracy of input data whose distributions varies from training. 
Ideally, the training and testing data should follow similar distributions, to avoid covariate shift, which means that the training dataset needs to cover a sufficient spectrum of application behaviors. Meanwhile, because of the impractical size of resource allocation space, Sinan's data collection agent is only able to cover a limited fraction of the entire space within the permitted time and computation budgets. As a result, we design the data collection agent to follow two principles. Firstly, we quantize the minimum amount of resources by which Sinan can adjust an allocation, to reduce the size of the explored resource space. 
% The much larger resource space of inter-dependent microservices makes random exploration prohibitively expensive and ineffectual. 
% For the \textit{Social Network} application specifically, collecting data every 1s for 24 hours only covers $4.45\times10^{-34}$ of the entire space, 
% even without considering that identical resource actions can result in different performance under different system states.
Secondly, we enforce the data collection agent to only explore allocations in the $[0, QoS + \alpha]$ tail latency region, where $\alpha$ is a small value compared to QoS, so that the trained model is able to learn the behavior of resource allocations which initiate QoS violations without biasing the collected distribution severely towards values greater than QoS. We have also compared Sinan's data collection scheme against collecting data randomly and when a resource autoscaler is in place, and showed that Sinan consistently explores a larger and more useful region of the resource space, improving the accuracy of the ML models. 
% Sinan's data collection explores resource allocations within the range of interest 
% of tail latencies. It constrains the space by turning the single constraint $\sum_{t} Core_t \leq Core_{total}$ to a set of 
% per-tier constraints $Core_t \leq C_t$, $t \in Tiers$, which reduces the allocation space from $\binom{C-1}{N-1} \cdot N^F$ 
% to $\prod_1^N C_i\cdot N^F$. The per-tier core constraints are intentionally over-provisioned 
% to account for maximum load and accelerate recovery from a QoS violation. 
%To further reduce the resource space, we also constrain the possible actions. 
%Fig.~\ref{fig:dist} shows the tail latency distribution of the training dataset for \texttt{Social Network}, 
%using our data collection, 

%\begin{wrapfigure}[10]{r}{0.23\textwidth}
%    \vspace{-0.3cm}
%    \centering 
%    \includegraphics[scale=0.3,viewport=60 20 400 315]{figs/dist.pdf}
%    \caption{Latency PDFs across data collection schemes.}
%    \label{fig:dist}
%    \vspace{-0.3cm}
%\end{wrapfigure}

\subsubsection{Resource allocation} 

Online scheduling in Sinan currently involves has two phases: core allocation and power management. In core allocation, the scheduler minimizes the number of cores until no further reduction is considered feasible by the ML model.
%If the model predicts 
% that the resource configuration can consistently meet QoS, the scheduler maintains the core allocation, 
% otherwise it will yield some cores back to the bottleneck 
% group of tiers to stabilize tail latency.
Then the scheduler enters the power management phase and gradually reduces frequency. 
%After several reductions, frequency either stabilizes by itself \textcircled{\footnotesize{6}}, 
%or after reverting some of the frequency reduction to compensate for the delayed queueing effect 
%\textcircled{\footnotesize{12}}, \textcircled{\footnotesize{10}}. 
After the two phases are complete, the scheduler keeps the resulting allocation, and increases resources when required by the ML model. %, and starts another round of scheduling.
The scheduler also has a safety mechanism for cases where the ML model fails to correctly predict a QoS violation. Whenever the number of missed QoS violations exceeds that threshold, 
the scheduler trusts the model less, and is more conservative when reclaiming resources. In practice, Sinan never had to lower its trust to the ML model. 
%whenever the scheduler spots that under a certain load level, the QoS violation frequency surpasses the threshold, the scheduler disables the model and keeps adding resources to all tiers until no more QoS violation is spotted, and records the final resource allocation. Later, the scheduler will always use the recorded resource configuration for the load until the machine learning models are retrained with the miss-predicted data or even more training samples.

\section{Evaluation}
\label{sec:Evaluation}
\vspace{-0.05in}

\subsection{Methodology}
\vspace{-0.05in}

We use a local cluster with 150 physical cores in total for data collection and online deployment. Each microservice runs in a Docker container. We collected 192,031 samples, and split them by 9:1 as training and testing set.

%{\color{red}- add specs of cluster. }

\vspace{-0.06in}
\subsection{Sinan's Accuracy and Speed}

We first compare the CNN in Sinan against a multilayer perceptron (MLP), and a long short-term memory (LSTM) network.
%, which is traditionally geared towards timeseries predictions. 
Sinan's CNN achieves the lowest RMSE (9.5ms vs. 19.6ms for MLP and 13.1ms for LSTM), while also having the smallest model size (264KB). Although the CNN's speed is slightly slower than the LSTM (6.7ms/batch vs. 3.6ms/batch for LSTM), its inference latency is within 1\% of the decision interval (1s), which does not delay online decisions. In terms of the BT model, validation accuracy is higher than 93\%, with 3.1\% false positives, and 3.9\% false negatives.
% , and a total of 
% 420 trees needed to predict performance evolution in the near future. 
In all cases, Sinan runs on a single NVidia Titan XP GPU with average utilization below 2\%. 
%It is also worth noting that the total training time for 300k samples using boosted trees is within 200s, which is an order of magnitude faster than using neural networks. {\color{red} you need to be careful with this. If boosted trees are that much faster why don't we use them for everything? }

\vspace{-0.05in}
\subsection{Online Deployment}
\label{sec:online_deployment}

We now evaluate Sinan's ability to meet QoS during online deployment. We compare Sinan against two autoscaling policies. AS\_Opt is configured according to~\cite{klimovic2018pocket}, which reduces cores and frequency when the CPU utilization of a tier drops below 30\% and 40\% respectively, and increases cores when utilization exceeds 70\%. AS\_Cons is more conservative, and optimizes for QoS. %It uses profiled thresholds to determine when resources should be scaled down and up. 
It uses 20\% and 30\% CPU utilization, to downsize cores and frequency respectively, and 50\% to upscale cores. For each service, we run 10 experiments with constant load from  10\% to 100\% of the max QPS, and a diurnal load pattern, where load starts from 10\%, gradually rises to peak QPS, and then decreases back to 10\%. 

%\begin{figure}[!t]
%    \centering
%    \begin{tabular}{@{}c@{}}
%    \includegraphics[scale=0.29,viewport=30 0 800 120]{figs/social_core.pdf}\\
%    \includegraphics[scale=0.29,viewport=30 0 800 115]{figs/social_freq.pdf}
%    \end{tabular}\\
%    \caption{Comparison of the average number of active cores and operational frequency between Sinan and Autoscaling. }
%    \label{fig:load_comparison}
%\vspace{-0.18in}
%\end{figure}

%\begin{wrapfigure}[10]{r}{0.18\textwidth}
%  \includegraphics[width=0.84\linewidth]{figs/violin_diurnal.pdf}
%\caption{Latency under diurnal load. }
%\label{fig:violin}
%\vspace{-0.18in}
%\end{wrapfigure}
At near-saturation load, differences between schedulers are small because of the limited size of our cpu pool. The difference becomes more apparent at low loads, where Sinan %is able 
reduces tail latency and latency variability considerably. In contrast, tail latency 
varies widely for the two autoscalers, and especially for AS\_Opt.
% , which optimistically downscales 
% allocations when utilization is 30-40\%. 
% Sinan reduces tail latency by 52.6\% and 11.1\% 
% on average compared to AS\_Opt and AS\_Cons, respectively. 
The violations in AS\_Opt are caused 
by not upscaling \texttt{NGINX}, whose utilization did not exceed the upscale threshold. 
%The difference is more dramatic for the diurnal load, shown in Fig.~\ref{fig:violin}, where AS\_Opt violates QoS by more than an order 
The difference is more dramatic for the diurnal load, where AS\_Opt violates QoS by more than an order of magnitude. %, due to utilization not being a good proxy for tail latency. 

Note that Sinan's tail latency reduction also comes with significant resource savings. 
Even when compared to AS\_Cons, Sinan reduces the active cores by \textbf{16.3\%} on average, and up to \textbf{29.1\%}. Sinan also reduces the average frequency of active cores by \textbf{37.2\%} on average, and up to \textbf{57.47\%}. 
%Compared to AS\_Opt, Sinan also significantly 
%reduces the QoS violation rate of the diurnal load from 8.3\% to 0.4\%. 

\begin{figure}
    \centering
    \begin{tabular}{@{}c@{}}
    \includegraphics[scale=0.09,viewport=1100 60 1650 420]{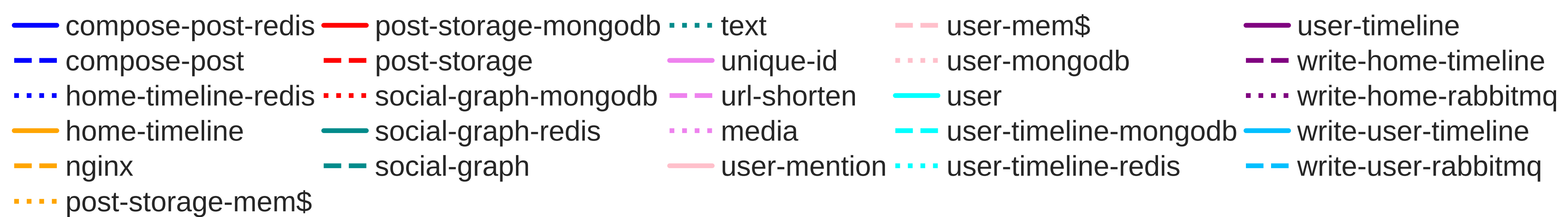}\\
    \includegraphics[scale=0.13,viewport=560 60 1400 455]{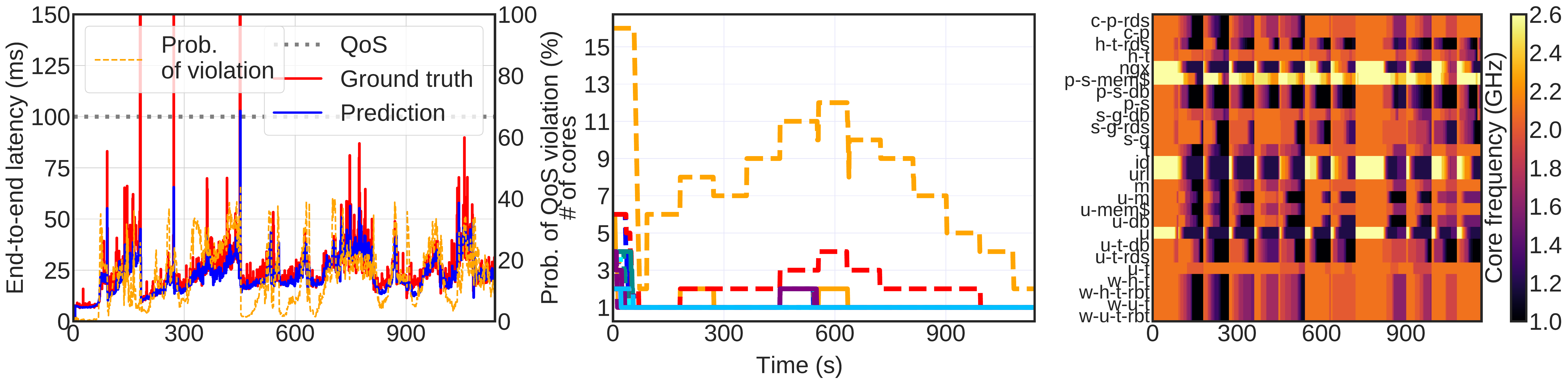}
    \end{tabular}\\
    \caption{\label{fig:detailed_deployment}Latency and resources under a diurnal load.}
   \vspace{-0.15in}
\end{figure}

Fig.~\ref{fig:detailed_deployment} shows the detailed results of Sinan's resource allocation over time for the diurnal pattern. 
% %We show the experiments for the diurnal pattern. 
% We show predicted vs. real latency and QoS violation probability, 
% the number of active cores, and core frequencies. %, and (c) the active core frequencies over time. 
Sinan is able to dynamically adjust resources to handle the fluctuating load, and the predicted tail latency closely follows the ground truth.

\vspace{-0.1in}
\subsection{Explainable ML}

For users to trust ML used in systems, it is important to interpret its output with respect 
to the system it manages, instead of treating ML as a black box. We are specifically 
interested in understanding what makes some features in the model %prediction process being
 more important than others. The benefits of understanding this %the importance of input features 
 are threefold: 1) debugging the ML models; 2) identifying and fixing performance issues; 
 3) filtering out insignificant features to reduce the model size and speed up inference.

%\subsubsection{Interpretability methods}
We adopt a widely-used ML interpretability approach called LIME~\cite{lime}. 
LIME interprets NNs by identifying key input features which contribute most to predictions.
Given an input $X$, LIME perturbs $X$ to obtain a set of artificial samples, close to $X$ in the feature space. Then, LIME labels the perturbed samples by classifying them with the NN, and uses the labeled data to fit a linear regression model, and uses it to identify important features based on the regression parameters. Since we are mainly interested 
in understanding the culprit of QoS violations, we choose input samples $X$ close to when QoS violations occur, and apply LIME. We perturb resource usage statistics, and construct 
a dataset with all perturbed and original data to train a linear regression model. 
%For example, to study the importance of MongoDB, 
%we multiply its utilization history with two constants 0.5 and 0.7, and generate multiple perturbed samples. 
Lastly, we rank the importance of each feature. % by summing the value of the associated weights.

%\subsubsection{Interpreting the CNN}
We applied LIME to diagnose performance issues in cases where tail latency experienced spikes despite the low load. First, %we identified the top-5 most important tiers 
%in terms of tail latency. We find that 
we find that the most important tier for the model's prediction is social-graph Redis, 
instead of tiers with heavy CPU utilization, like \texttt{NGINX}.
We then examine the importance of each resource metric for Redis, and find that the most meaningful resources are cache and resident working set size, which correspond to data cached in memory and non-cached memory for a process, including stacks and heaps. 
Using these hints, we check the memory-related configuration and runtime statistics 
of social-graph Redis, and find that it is configured to record logging data in persistent storage every minute. For each persistence operation, Redis forks a new process and copies all written memory to disk; during that time it stops serving user requests. Disabling logging resulted in most latency spikes being eliminated. Re-applying LIME to the modified Social Network showed that the importance of Redis had significantly dropped, in agreement with our observation that tail latency is no longer sensitive to it. %that tier. %'s operation.

\section{Conclusion}
\label{sec:Conclusions}

We have presented Sinan, a scalable and QoS-aware resource manager for interactive microservices. 
Sinan highlights the challenges of managing complex microservices, and leverages a set of 
validated ML models to infer the impact allocations have on end-to-end tail latency. 
Sinan operates online and adjusts its decisions to account for application changes. 
We have evaluated Sinan both on local clusters and public clouds GCE) 
across different microservices, %over different end-to-end applications built with microservices, 
and showed that it meets QoS without sacrificing resource efficiency. Sinan highlights 
the importance of automated, data-driven approaches that manage the cloud's complexity in a practical way.

\section*{Acknowledgements}

We sincerely thank Daniel Sanchez and the anonymous reviewers for their feedback on earlier versions of this manuscript.
This work was in part supported by NSF grants NeTS CSR-1704742, CCF-1846046, and a John and Norma Balen Sesquisentennial Faculty Fellowship.

%\clearpage
\balance
%

% The next two lines define the bibliography style to be used, and the bibliography file.
\bibliographystyle{IEEEtranS}
\bibliography{references}

%\newpage
%\appendix
%\input{Appendix.tex}

\end{document}